# Cultural fragmentation and innovation diffusion in a dynamic scenario

*Andrea Apolloni[1], Floriana Gargiulo[2]*

[1] *Network Dynamics and Simulation Science Laboratory (N.D.S.S.L.),
Virginia Tech, Blacksburg (VA), USA*
[2] *,Laboratoire pour l'Ingeniere pour les System Complexes (LISC),
Cemagref, Clermont Ferrand, France*

## ABSTRACT

Axelrod's model describes the dissemination of a set of cultural traits in a society constituted by individual agents. In a social context, nevertheless, individual choices toward a specific attitude are also at the basis of the formation of communities, groups, parties. The membership in a group changes completely the behaviour of single agents who start acting according to a social identity. Groups act and interact among them as single entities, but still conserve an internal dynamics.
We show that, under certain conditions on social dynamics, the introduction of group dynamics in a cultural dissemination process avoids the flattening of the culture on a single thought and preserve the multiplicity of cultural attitudes.
We also considered an innovation diffusion process on this dynamical background, showing under which conditions innovative ideas can survive in a scenario where the groups' choices determine the social structure.

## INTRODUCTION

The processes that give rise to the formation of new *jamming cultures* or to the imposition of predominant ways of thinking are important objects of analysis in the globalization era.
While various examples of integration appear as an effect of the migration flows and of the extension of the social networks to the world level, simultaneously phenomena of racism and conservative attitudes are more and more evident in every part of the world.
The cultural dissemination processes are spreading mechanisms in which, through imitative attitude, each individual tends to adopt and replicate the social norms that are predominant in its environment.
When people with completely different cultural backgrounds enter in contact, a new equilibrium with a single jamming culture can be reached, or, under different circumstances, cultures do not integrate with each other and cultural barriers are erected.
The diffusion of cultural norms in a static society is described by "Axelrod's model", (Axelrod 1997).

Another phenomenon that is attracting the attention of social scientists is the increasing participation in social networks, above all those introduced with the WWW (Facebook, Flickr, etc.). This kind of tool provides enormous quantity of data and allows analysis of social groups behaviour that, with the methods of the traditional sociological research, were unthinkable.

Individual choices toward a determined cultural attitude are the basic ingredient for the formation of communities, groups and parties. Groups promote homophile relationships between individuals who are similar and increase the cultural barriers to the dissimilar ones (Popielarz 1995).
Groups act and interact with each other as single entities, compete for membership recruitment and represent the common feeling of its members toward a determinate norm.
The membership in a group changes completely the behaviour of single individuals that are inside the group: they start acting according to a social identity (Tajfel 1986).

The study of social groups and their dynamics plays an important role in mining and analysing phenomena based on social data (Backstorm 2006): evolution of informal groups within a large organization could give some insight in decision making; the spreading of diseases strongly depends on society stratification and dynamics (Morris 1993); market dynamics depend on herd behaviour of its agents (Eguiliz 2000); firm growth depends on economical behaviour of firms (Schweitzer 2008).
In recent years, efforts in studying group dynamics have been directed along two directions: Identifying groups in large networks (Girvan 2002; Fortunato 2007); studying the mechanisms that lead to formation and evolution of groups, identifying the structural elements which determine the membership properties, the size and the natural decline (Backstorm 2006). The emergence of social groups as an effect of the homophily dynamics was described in (Centola 2007).
Related to the latter approach is the study of how group size distribution emerges from the individual behaviours. Two fundamental dynamics have been identified for the evolution of groups (Johnson 2006; Levin 1995; Ruszczycki 2008; Zhao 2009): coalescence (two groups merge together) and fragmentation (a single group can split in smaller groups as in the case of Zachary karate club (Zachary 1977) ).
These dynamics have been already studied in different contexts, such as physics and chemistry, where interaction is considered only between nearest neighbours (particles and molecules) and the detailed energetics of the microscopic reactions leads to formation of macroscopic patterns. Instead in social systems the precise microscopic rules are less evident and the formation itself of groups (corresponding to macroscopic patterns) in a social environment could in many cases be a by-product of the process. Moreover due to the introduction of new communication technologies, interaction among agents is not limited to nearest spatial neighbours and the very definition of neighbours needs to be re-formulated.
Strictly related to group dynamics are diffusion processes. It has been pointed out by Morris (Morris 1993) that the spreading of a disease like HIV/AIDS can not really be understood/simulated if the interaction among groups, in particular mixing processes, are neglected: mixing among groups can get the prevalence lower if the prevalence is low in a group, increase in the opposite case.
Of particular interest is the experimental setup described in (Barrat 2008) to study this kind of dynamics.
The objective of this article is twofold. We present a model of group dynamics similar to (Johnson 2006; Levin 1995; Ruszczycki 2008), where the coalescence of groups is ruled by a modified homophily interaction. Parallel to group dynamics we implant a diffusion of innovation process. In our model the innovation diffusion process is seen as an epidemic spreading model where the new idea is transmitted with a certain rate inside the neighbourhood.
In our model we consider also a conservative movement to the introduction of the novelty. The resulting model then consists of the contemporaneous diffusion of these two behaviours.

## CULTURAL TRAITS AND SOCIAL GROUPS

In the seminal work by Levin (Levin 1995) and in more recent papers, a model for group dynamics based on coalescence and fragmentation has been proposed (Johnson 2006; Zhao 2009).

In all these cases the authors consider a population divided into a certain number of groups. At each time step a group is chosen, and it could either split with a probability proportional to its size or merge with another randomly chosen group. The details of the fragmentation dynamics differentiate the two models.

Whilst the process of coalescence is the same for groups of the same party, the interaction between groups of different parties could not occur (Levin 1995) or lead to the destruction of the smaller one and an equivalent reduction of in the size of the bigger (Zhao 2009).

In all these cases groups are just characterized by size and the coalescence process doesn't take account of any detailed mechanism of merging. Instead individuals, and then groups, try to interact with people who are perceived as "similar", where the similarity is based on the characteristics which determine the membership to a particular group.

People tend to characterize themselves with social labels: races, political views, religious beliefs. Using these labels they declare their membership in particular groups: people identify their "social DNA" and interact preferentially with people who share many common traits.

Once they form a group a social identity is created: they act as a collective body, characterized on its time from a collective social DNA.

To study the society means also to describe the coarsened community structure that is its structural backbone: parties, religious sects, online communities present universal behaviours independent from their specific social environment.

Groups feel the distance from other groups and do not interact with whom they perceive as too distant (Tajfel 1986). But, on the other side, in order to acquire social prestige and to avoid isolation, they try to enlarge themselves by acquiring new members (Popielarz 1995), or joining with similar groups . The concept of distance itself is relative to the open-mindedness of the groups: two groups can consider themselves distant if they differ for just one cultural trait (close mind) or close if they have no common trait (open mind). At the same time, when a group reaches a critical size, internal discrepancies can compromise the stability of the group and can lead to the birth of a new group: a sort of social mitosis where a mutation of the original group is originated and becomes independent.

Looking at the society at the scale of social groups we can notice that social groups have a precise life cycle: they are born with a mutation process from a pre-existing group, they evolve trying to assimilate new members and they die if they are absorbed into a bigger community.

Also in the historical formation of the National States these aggregation and division processes can be somehow identified: phases of assimilation and of homogenization are continuously alternated with phases of strong nationalism, with clear splitting tendencies (Cederman 2006).

At a larger scale, if we consider the social system in its totality, the lifetime of a particular cultural trait is different from the lifetime of a single group. A trait that dies with a group can survive or be reborn in a different community.

In our model we define a group as a clique of at least 3 agents, sharing the same string of cultural traits, namely the social attitude.

We identify the social DNA of a group as a string of binary bits of length L; in a society, we can identify at most $2^L$ different cultural identities.

Every group is characterized by its social string and by the number of its members, $n_i$.

We can also have simultaneously groups with the same DNA structure, that for their historical pattern or for any other reason (i.e. geographical distance and isolation) cannot converge into a single entity.

The cultural similarity between two groups can be measured by the overlap function (Centola 2007):

$$\Theta(i,j) = \sum_{h=1}^{L} \delta_{\phi_{ih}\phi_{jh}} \qquad (1)$$

This function is ranging between zero and L: it assumes the value zero when the two cultural strings do not have nothing in common and L when they are coincident.

The dynamics of the group can allow two different processes: coalescence and fragmentation. We characterize these two different tendencies with two parameters, ε and α.

Two groups can join if their overlap function differs for less than a certain threshold value.

If two communities join, the majority group imposes its cultural traits on the smaller one so that the latter one dies in the process.

Simultaneously each group can present a certain tendency to fragment that is proportional to its number of members:

$$p_{frag} = \alpha * (\text{size of the group})/N \qquad (2)$$

Where N is the total number of agents involved in the simulation.

If the group divides, a new group is generated, composed of a random percentage of the members of the begetting group. Its social string is just the mutation of one trait from the social DNA of the begetting group.

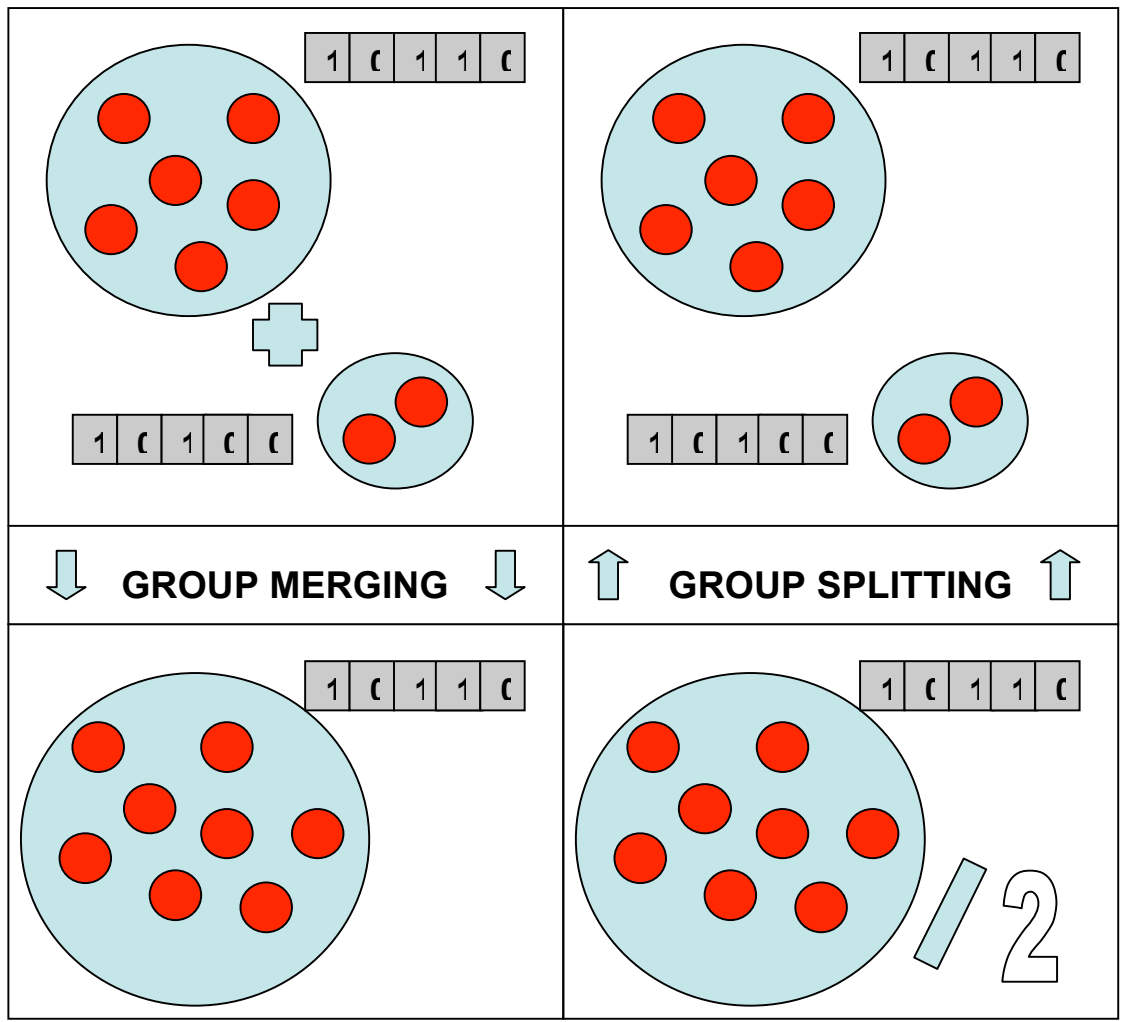

**Figure 1**: A visual review of the two possible types of group interaction

In the simulation process we randomly chose phases of coalescence and fragmentation of groups: at each step a group randomly decides which process it undergoes. If it decides to join to another group, it chooses randomly the second group and if they are compatible according to the threshold they merge into a new group. If it decides to split, it does with probability $P_{frag}$ generating a new group.

At the level of the social network, all the agents inside a group constitute a complete graph; each group could be in principle connected with all the other groups, but, in fact they interact only with groups with a cultural string differing less than a fixed threshold. Therefore all the agents in a specific group can only interact and be connected with agents sharing similar ideas inside or outside the same group.

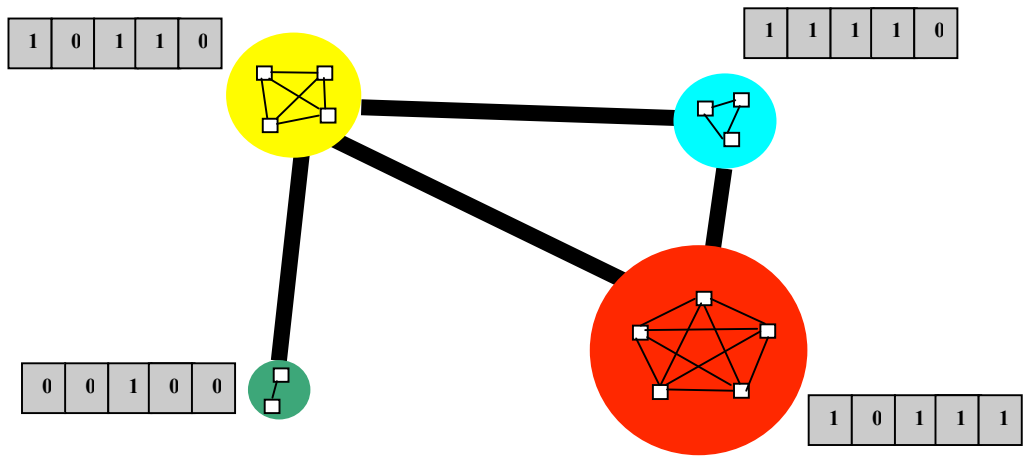

**Figure 2:** The structure of the social network. Black lines indicate some of the possible interactions among groups

# DIFFUSION OF INNOVATIONS

Getting a new idea adopted, although representing a crystalline vantage, is a process lengthy, complex and many times also unsuccessful. Studying how to speed up this process means taking account of different factors such as time, newness and social characteristics.

An example is the in-field case of introduction of hot water practice in Peru', reported in the first pages of Rogers, "diffusion of innovations" (Rogers 2003; Valente 2005): as a part of Peru's health agency program to introduce boiled water in daily routine, a promoter agent is sent in the village "Las Molinas" to persuade housewives. The campaign is unsuccessful. In the analysis of the campaign important factors emerge: the health agency did not take into account the local beliefs about hot food and drink; the only adopters were village outsiders.

The above example makes explicit some dynamics of diffusion processes. A diffusion process is a communication of a new idea by an innovator to the rest of the community, which needs a social system for the transmission and that could be related to a social change.

The communication is facilitated if both the innovator and the rest of the community share common social/cultural traits. The probability of adopting certain behaviour is influenced by the number of people who have already accepted, i.e. social exposure, and as demonstrated, also in the seminal work by Gross and Ryan (Valente 2005), inter personal communications and imitation processes are the best means for conveying idea.

An innovator is an individual more exposed to mass media and/or endowed with cosmopolitan attitude, often seen as a deviant with respect to society norms. The role of innovator is then limited and depends on the presence of opinion leaders. If a society is progressive and oriented to change (new social attitudes can be created or old ineffective ones destroyed) opinion leaders are innovative, whilst in the opposite case they reflect social norms and are opposed to change.

Bass has summarized the above dynamics in a mathematical model (Bass 1969). Both in the original formulation and in his stochastic version (Isii 1959; Shun-Chen 2002), agents in the population are divided into different classes: innovators, namely the first adopters of the new behaviour, and adopters/imitators, namely individuals that under peer pressure of innovators, opinion leaders or other adopters adopt the new idea. Bass model depends on two parameters: the fraction of innovators, $p$, and the probability of transmission of new behaviour due to a contact between an adopter and another member of the population, $q$.

The time evolution profile of the number of the adopters is s-shaped: after a slow increase the adopters reach a critical mass and then the process exponentially grows until reaching a stable value until a newer innovation is introduced.

The Bass model can easily be mapped in an epidemiological model SI, (susceptible infected) where the innovator is the seed of an epidemic and the parameter $q$ is the probability of infection due to a contact between an infected and a susceptible individual.

Bass model implicitly admits that propagation of innovation is mainly due to interpersonal relations and word of mouth; moreover it describes the hierarchical spreading before in the group and later outside the group.

Bass model, on the other side, has a big limit: it excludes the possibility of coming back to the previous attitude. In the Bass model only the innovator can influence the rest of society, and once it happens the agent cannot go back from such a state. Instead communication is two-way and each interlocutor tries to convince the other of its position. In societies censorship systems exist in reaction to the birth of a new idea or social norm: adopters and the innovators themselves can recover from their status (i.e. G.Galilei). The diffusion process can be seen as a tug of war between innovators and adopters on one side and resistance on the other. In this view, the innovation diffusion can be described as a dynamical competition between two species of ideas, the innovative one (A) and the conservative one (B). In such competition both the transitions are allowed.

An approach that takes into account both the possibility of the innovator and of the conservative to convince each other is the Abrams-Strogatz model for language competition (Abrams 2003; Stauffer 2007). The studied case for Abrams-Strogatz model regards the progressive affirmation of a unique language in a mixed population initially speaking two different idioms. The adoption of a language depends both on the number of individuals that already adopt the language and from the prestige of the idiom.

We use such approach for describing the diffusion of an innovative idea A in conservative population with idea B.

People can decide to adopt an idea (A or B) for direct imitation and according to the number of persons that have already adopted the idea.

A transmission rate $\beta_A$ is associated to the transition from the conservative to the innovative idea:

$$B+A \rightarrow 2A$$

At the same time conservative people offer resistance to the introduction of innovation: conservatives try to convince innovators to go back to the original idea (B). The transmission rate in the case

$$A+B \rightarrow 2B$$

is $\beta_B$.

The normalized number of adopters of the idea A at the time $t+1$ ($a_{t+1}$), assuming that every individual can interact with the rest of the population, is:

$$a_{t+1} = a_t - (p_{A \to B}(a_t))a_t + (p_{B \to A}(a_t))(1 - a_t) \qquad (3)$$

The second term of the right member of the equation represents the individuals that at each time step transit from A to B status. The third one is the opposite transition.
The probability transitions $p_{A \to B}$ and $p_{B \to A}$ depend both on the transition rate of the reaction and on the precedent abundance of the resulting idea of the transition:

$$p_{A \to B} = \beta_B (1 - a_t)^\lambda$$
$$p_{B \to A} = \beta_A (a_t)^\lambda \qquad (4)$$

The factor $\lambda$ represents how important is the fact that the idea is already supported by a large number of persons. In the following we will consider the case $\lambda=1$.
Performing the change of variable:

$$x_t = \frac{(\beta_A - \beta_B)}{[1 + (\beta_A - \beta_B)]} a_t \qquad (5)$$

and substituting equations (4) in equation (3) the map for the evolution of $x$, finally results to be:

$$x_{t+1} = [1 + (\beta_A - \beta_B)] x_t (1 - x_t) \qquad (6)$$

This is the logistic map (Ott 2002) with parameter $r = [1 + (\beta_A - \beta_B)]$.
Since $\beta_A$ and $\beta_B$ vary in the range [0,1], the parameter $r$ of the logistic map can be either in the range $0<r<1$ or $1<r<2$, according to the values of $\beta_A$ and $\beta_B$: if the strength of the innovative idea is larger than the strength of the conservative one ($\beta_A > \beta_B$) the parameter $r$ is in the range $1<r<2$, and consequently the fixed point is:

$$\tilde{a}_{(\beta_A > \beta_B)} = 1 \qquad (7)$$

Namely, the innovative idea spreads to all the population.
In the opposite case, if the conservative idea is stronger than the innovative one, $r$ is in the range $0<r<1$ and consequently the fixed point is:

$$\tilde{a}_{(\beta_B > \beta_A)} = 0 \qquad (8)$$

corresponding to the death of the innovative idea.

Also the division of the society in different groups is important for the spreading of an innovative idea.
In order to have diffusion, a certain level of heterogeneity between agents should be present (Rogers 2003). In the previous section we have defined group based on its social DNA or social attitude. From this point of view the communication within a group is facilitated.
But while homophily can speed up the process in the early stage, diffusion on the social network can take place only if a critical mass of adopters is reached and then the reaction is self-sustained. The homophily in this way represents an obstacle to diffusion: the communication is restricted just to individuals belonging to the same clique. In order to spread widely the idea, in the social network some links should be present which connect different cliques of individuals. The extreme examples of the importance of this kind of "heterophilus" links can be found in Granovetter theory of strength of weak ties (Granovetter 1973), and in some caste based societies (where the lack of connection between castes limit the spread of innovations) (Rogers 2003).

In our model, the introduction of a new idea is related to a cultural trait that is not fundamental for group formation, this means that two individuals in the group have identical cultural traits but express a different opinion on the novelty. On the other hand, from an epidemiological point of view, innovation can be seen as a virus transmitted independently of cultural traits, so individuals belonging to different groups are as susceptible to change as people in the same group. The final outbreak size, or the final number of adopters depends only on the underlying network, not on the particular social attitude of the group (or its members). The new opinion first diffuses within the group and then, due to the coalescence and fragmentation dynamics and its consequent repartition of members, can be transmitted to other groups.

At the beginning of the diffusion there is just one innovator surrounded by a sea of resistant people; the number of people with whom the innovator is in contact depends on the initial partition in groups. We could associate with the innovator a larger probability of infection, reflecting a higher leadership, newness of the idea and impact.
To include the group structure inside the innovation diffusion model and to consider stochastic oscillations, instead of the approach with the logistic map described previously we used a binomial extraction process inside each single group. This kind of approach is similar to the metapopulation epidemic model proposed in (Colizza 2006).
We consider that the contamination process of an idea takes place between agents that are members of the same group. At each time step, therefore, for each group *i* a double "contamination" mechanism is considered: each fraction of the group tries to convince its opponent.
Consider for example the number of individuals with idea A, in the *i*-th group $A_i$, and the complementary number of individuals with idea B, $B_i = n_i - A_i$. At each time step $A_i$ can increase due to the adoption of the idea A by some of $B_i$, or could decrease due to the adoption of idea B by some of its member.
In any case the adoption probability depends on both the transmission rate of the idea and the number of agents that already support it.
For the first process (increase of $A_i$), in the assumption of homogeneous mixing inside the group, the probability rate is given $\beta_A A_i(t)/n_i(t)$, for the second one by $\beta_B B_i(t)/n_i(t)$. The number of new adopters of each idea is then a stochastic variable that follows a binomial distribution with the corresponding adoption probability. The net change of adopter of a specific idea, at time step *t*, is given by the difference between these variables.

## SIMULATION APPROACH AND RESULTS

In the following are described the details of the model, the analyzed quantities and the statistical analysis performed on the results.
We consider a population of size N=2000 initially divided in NC groups. NC can assume different values in the range between 1 and 100, describing extreme situations: a unique original group as starting point or an already highly fragmented society.
The agents are randomly assigned to each group at the beginning of the simulation.
We indicate with $n_i$ the size of the i-th group.
With each group is associated a social attitude which is represented by a vector ϕ, with binary entries and of length L.
At every time step a group *i* is randomly chosen and subjected to one of the two possible actions:
1) Fragmentation: With probability $p_{frag}=\alpha n_i/N$, i.e. proportional to the size of the group, the group, due to internal reasons, can split up and form two subgroups. The smaller ones is different from the group mother, changing one of the characters of its social attitude. The quantity α is a parameter of the model called the splitting rate. We consider only two extreme cases: a static population, α=0, and a dynamical one, α =1. In the first case the fragmentation process is completely suppressed and the creation of new cultures during the simulation, different from the initial ones, is forbidden.

2) Coalescence: Another group *j* is randomly chosen and the two social attitudes $\phi_i$ $\phi_j$ are compared using the overlap function (1). If the overlap function is smaller than (εL), namely if the two vectors differ by less than (εL) elements, the two groups merge together. The social attitude of the group is given by majority rule: the cultural traits of the biggest group are conserved. The parameter ε is called open-mindedness and in the simulation it's varied from 0=closed mind (groups never merge) to 1= extreme openness (independently of social attitudes two groups always merge) in step of 1/L. In our numerical simulations L is set equal to 5.

For all the possible values of ε we evolved the system for a time T=500, expressed as iterations. To deal with the intrinsic stochasticity of the system, the experiment has been repeated 500 times. The various measures reported in the graphs are obtained as an average of the 500 realizations.

Several measures have been performed. At each time step we have computed the number of different groups and the number of independent social attitudes, contemporary with the measure of the size of the biggest group and biggest social attitude.

A social attitude can be shared by different groups; therefore it could happen that a condition of majority (bigger social attiude) could correspond to a situation where several small groups, geographically or historically distant, are simultaneously members. In this context the merging of a small group of the majority with a bigger one of a different attitude, could change the balance of power. This is particular true in a bi-partite system where one of the political currents is slightly bigger than the opponent, but highly fragmented in small parties: in such cases one of small parties of the opposition can decide to merge with the opposition, changing the majority.

Simultaneously to the network dynamics also a process of innovation diffusion is considered using the process described in the previous section.

In our model we have consider just one initial innovator located into a random group.

The parameters are fixed to $\beta_A$=0.6, $\beta_B$ =0.2, which simulates at the same time a high impact factor for the new idea and also a strong resistance from the uninfected people The seed can initially infect only people belonging to its same group. The innovation can spread among the other groups only if a group containing at least one innovator merges with another one.

If a group is completely infected or completely susceptible, it cannot recover spontaneously unless it merges with another group that is not completely infected.

For describing the innovation diffusion mechanism, we have evaluated at each time the number of people adopting the innovative idea (prevalence of innovation). At the same time we have reported the final size of the contamination as a function of ε

The simulation describes three independent levels of dynamics: the dynamics of groups, the dynamics of social attitude (majority composition) and the dynamics of the innovation propagation on this dynamical network structure.

These different levels of analysis are described in different paragraphs.

We have to analyze the effects of three different factors characterized by three parameters:
- Tendency to create fragmentation, parameterized by α
- Open mindedness of the groups, parameterized by ε
- Initial fragmentation, parameterized by NC

In the following sections the results of six sets of experiments are reported, for all the possible values of ε:

EXP1: α=0, NC=0
EXP2: α=0, NC=50

EXP3: α=0, NC=100
EXP4: α=1, NC=1
EXP5: α=1, NC=50
EXP6: α=1, NC=100

## Group Dynamics

In Figure 3-7 are displayed the results concerning the analysis of the group dynamics for the various experiments introduced in the previous section. In particular, we analyze the number of groups at each step of the simulation, the size of the biggest community, and the size distribution of the groups at the final state. The different colours refer to different values of the open-mindedness parameter.
The different parameter setup can give rise to completely different final configurations..
The results for EXP1 are not reported since in this case the dynamics are trivial: starting from one group (NC=1) and excluding the possibility of creating new groups (α=0), it will remain a single community during all the simulation.
Figure 3 and 4 refer to the other two experiments with α=0 (EXP2, EXP3), namely where the creation of new groups is suppressed. For α=0, the dynamics for ε=0 (black line) are trivial: in this case, groups never merge, even if extremely similar in the social attitudes, killing the dynamics.
For higher values of ε, three different regimes can be identified:

- For ε=1/5 (red line) the number of groups decrease to a fixed value depending on the initial fragmentation (NC). In both experiments, at the end of the simulation, a reduction of the 50% of the number of groups is reached. The case ε=1/5 is examined in detail in Figure 8. In this Figure the number of final communities is plotted as a function of the initial number of communities NC. For an elevated level of initial fragmentation (NC>=50), $N_{final}$/NC is fixed to 0.5.
    
    The right plots of figure 3 and 4 show that the size of the various communities can vary continuously between a very small size (<5 agents) until containing half of the agents. The biggest group (inbox of the left graph) contains around the 35% of the population. The groups at the final state are extremely heterogeneous for the size.
- For ε=2/5 (green line) the situation is different. In this case a small number of groups are maintained at the end of the simulation: between 2 and 3 groups for EXP2 and between 3 and 4 for EXP3. From the left plot and from the inbox we can argue that a giant community (containing on average 80% of the population) is formed. The remaining groups, therefore, turn out to be extremely small.
- For ε>2/5 (blue line), the open-mindedness allows the final convergence into a unique group.

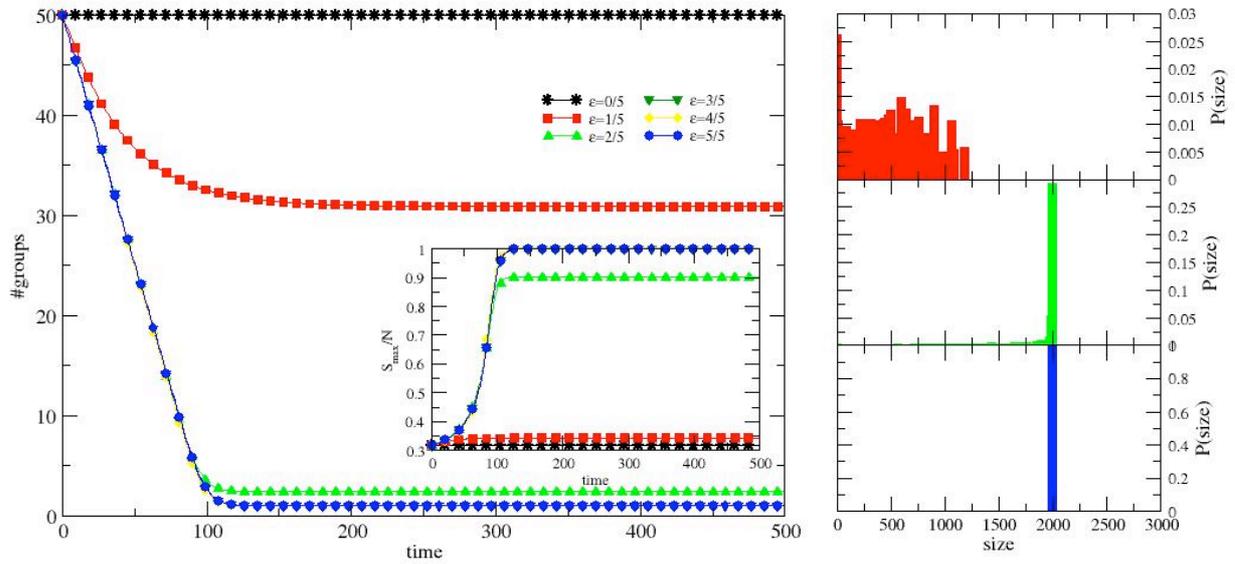

**Figure 3**: Left plot: The number of groups as a function of time for different values of ε in the case of EXP2. Inbox: Normalized size of the biggest group as a function of time. Right plot: Size distribution of the groups at the final state, for ε=1/5, 2/5, 5/5. The results are averaged over 500 realizations of the system.

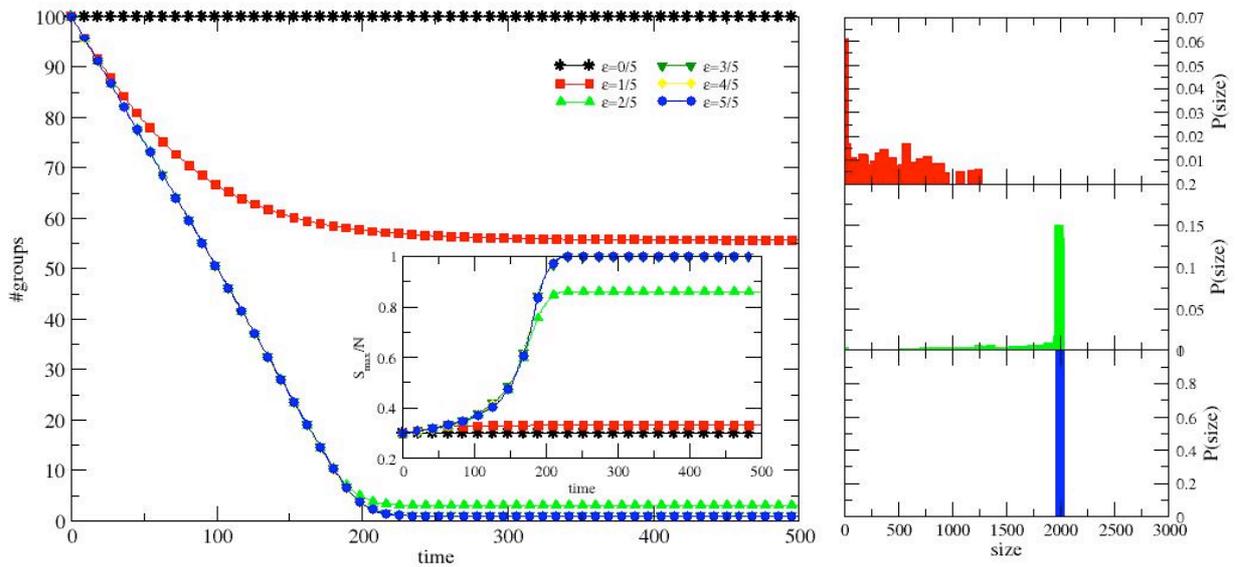

**Figure 4**: Left plot: The number of groups as a function of time for different values of ε in the case of EXP3. Inbox: Normalized size of the biggest group as a function of time. Right plot: Size distribution of the groups at the final state, for ε=1/5, 2/5, 5/5. The results are averaged over 500 realizations of the system.

Figure 5-7 refers to the three experiments with α=1 (EXP4, EXP5, EXP6). Also in this case we can identify three different regimes, according to the possible values of ε.

- For ε=0, independently from the initial fragmentation the number of groups can only increase during the simulation. The final state presents an extremely elevated number of groups of small size.
- For ε=1/5, exactly as in the cases with α=0, the final number of groups stabilizes around a fixed value. As we can observe from figure 8, for NC<30 the final number of groups is higher than the initial one, while for NC>30 the final number of groups is lower than the initial one. As in the cases with α=0, for an elevated initial fragmentation (NC>=50) the number of groups at the end of the simulation is reduced to 50% with respect to the initial situation.
  The size distribution of the groups at the final state is very different in these cases: the group sizes are more homogeneous since they are distributed with a small deviation around an average value that is:
  
  AV=266.6 for NC=1
  AV=155.85 for NC=50
  AV=147.7 for NC=100
- For ε>1/5 it is realized the same situation observed in the cases with α=0 and ε>2/5.

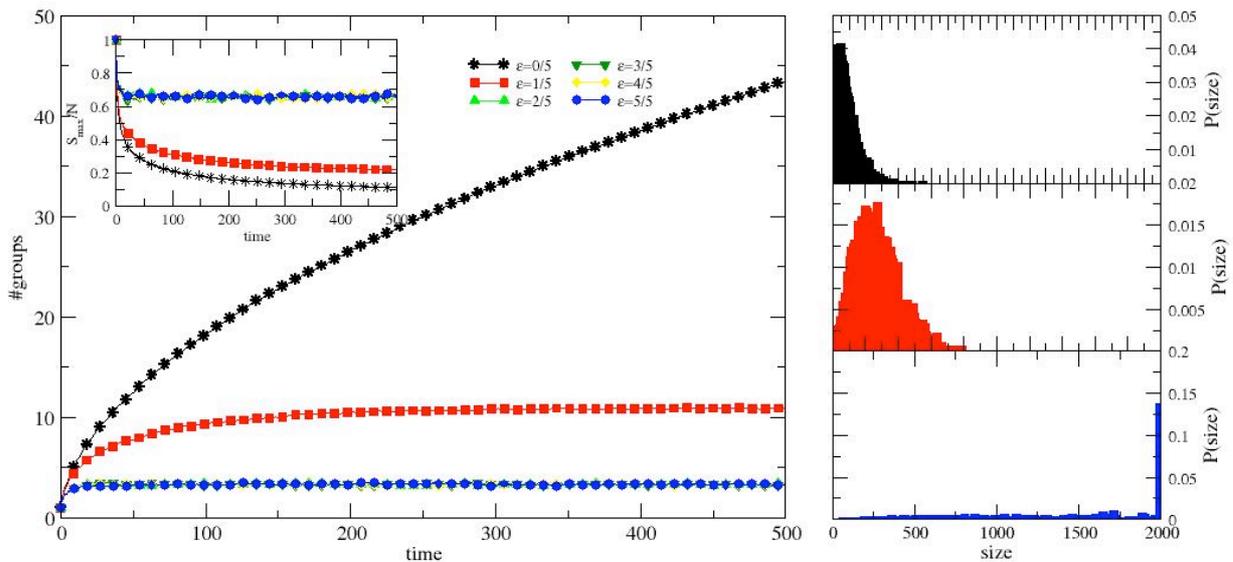

**Figure 5**: Left plot: The number of groups as a function of time for different values of ε in the case of EXP4. Inbox: Normalized size of the biggest group as a function of time. Right plot: Size distribution of the groups at the final state, for ε=0/5, 1/5, 5/5. The results are averaged over 500 realizations of the system.

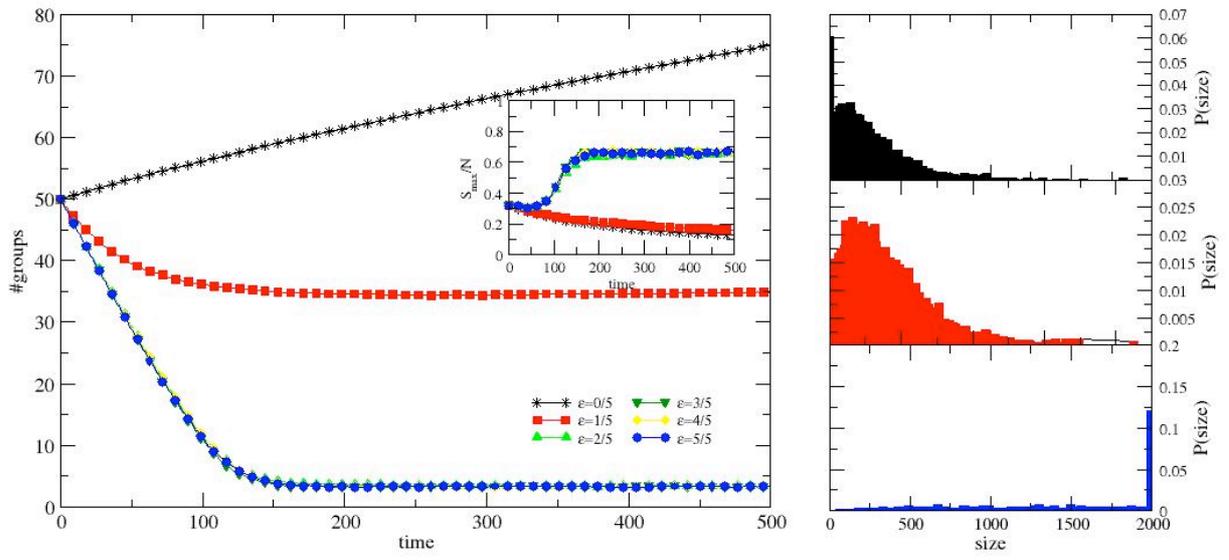

**Figure 6**: Left plot: The number of groups as a function of time for different values of ε in the case of EXP5. Inbox: Normalized size of the biggest group as a function of time. Right plot: Size distribution of the groups at the final state, for ε=0/5, 1/5, 5/5. The results are averaged over 500 realizations of the system.

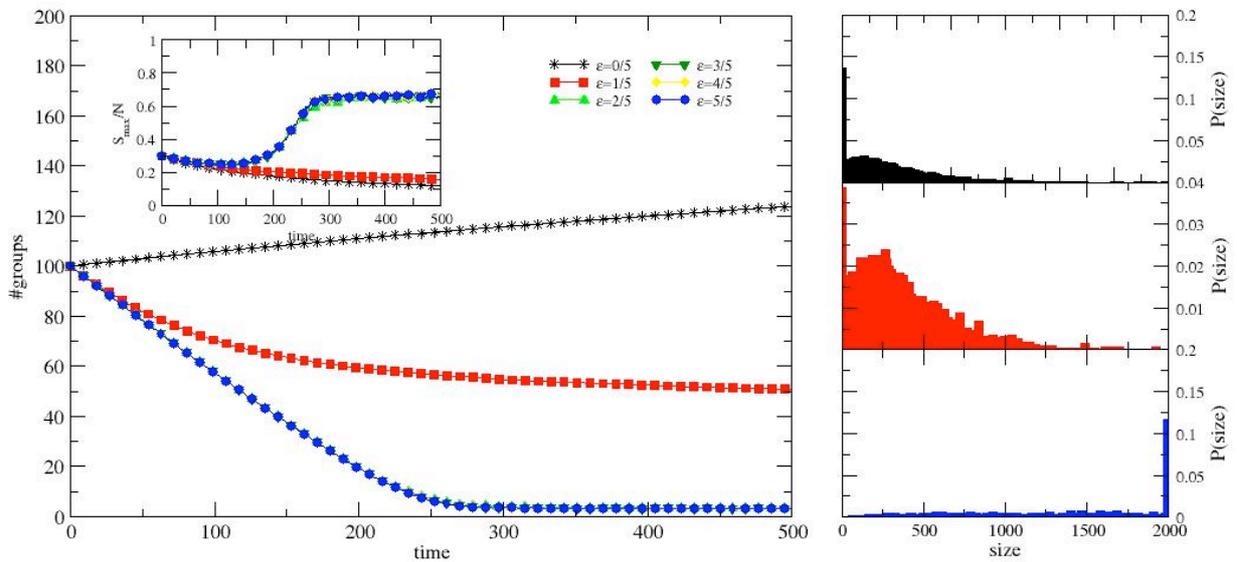

**Figure 7**: Left plot: The number of groups as a function of time for different values of ε in the case of EXP6. Inbox: Normalized size of the biggest group as a function of time. Right plot: Size distribution of the groups at the final state, for ε=0/5, 1/5, 5/5. The results are averaged over 500 realizations of the system.

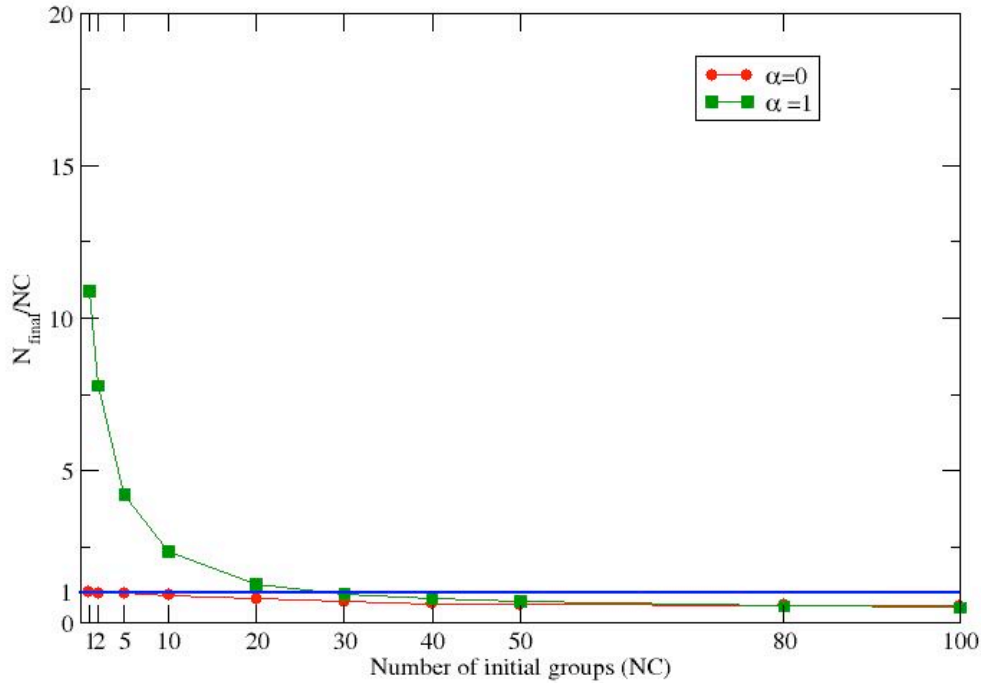

**Figure 8**: Number of final groups divided by the number of initial groups as a function of the number of initial groups for ε=1/5.

## Social Attitude Dynamics

Figures 9-13 display the results related to the evolution of independent social attitudes.
The trivial case EXP1 is not reported: due to the absence of any dynamics there is only one social attitude. In the following for size of a social attitude is intended the number of agents having a certain social attitude irrespectively of the group they belong to: a small social attitude means a social attitude shared by a small fraction of the agents, viceversa for large.
Figures 9 and 10 refer to the other two experiments for α=0 (EXP2, EXP3). As in the case of groups' dynamics we can distinguish three regimes:
1) ε <2/5 The number of independent social attitudes remains constant, independently on the initial distribution of social attitudes. At the end of the process the society is highly fragmented.
2) ε =2/5 Society is fragmented. In the case of initial fragmentation NC=50, Figure 9, society is almost tripartite since the number of independent social attitudes at the end of the process is 2.5, indicating the equal possibility for being tri or bipartite. Instead in case of initial fragmentation NC=100, Figure 10, society is tripartite. In both the case there is a large size social attitude and one or two small social attitudes, as shown in the inset and right plots. Compare to the previous regime the probability of having a large component is non null.
3) ε > 2/5 Consensus is reached: there is a unique social attitude in all the society. This corresponds to the fact that all the groups have already merged together.

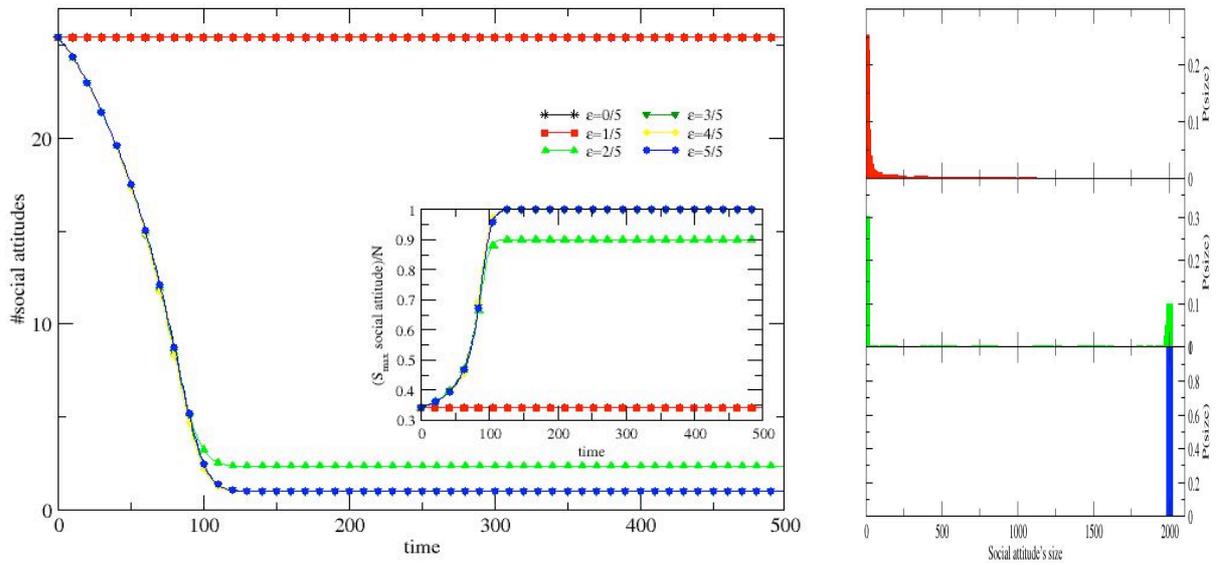

**Figure 9:** Left plot: The number of independent social attitudes as a function of time for different values of ε in the case of EXP2. Inbox: Normalized size of the biggest social attitudes as a function of time. Right plot: Size distribution of the social attitudes at the final state, for ε=1/5, 2/5, 5/5. The results are averaged over 500 realizations of the system.

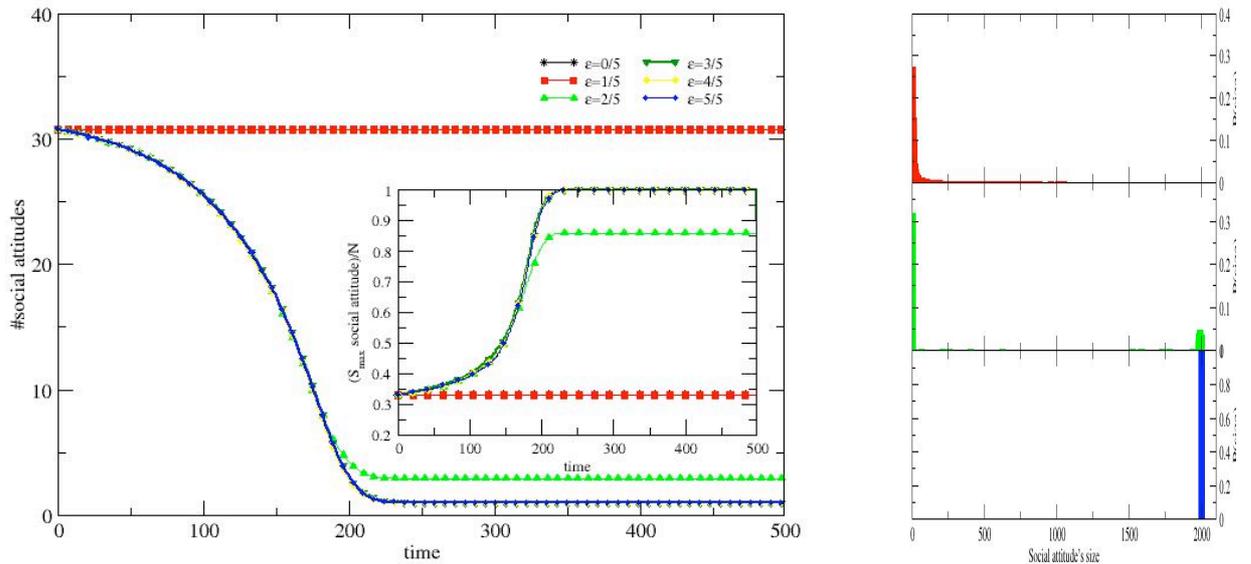

**Figure 10:** Left plot: The number of independent social attitudes as a function of time for different values of ε in the case of EXP3. Inbox: Normalized size of the biggest social attitudes as a function of time. Right plot: Size distribution of the social attitudes at the final state, for ε=1/5, 2/5, 5/5. The results are averaged over 500 realizations of the system.

In a dynamical population, instead, the social attitude dynamics is different as shown by Figures 11-13, referring to experimental sets EXP4, EXP5, EXP6. Also in this case we can distinguish three different regimes for ε, although the values are different.

1) ε = 0 The number of independent social attitudes increases This is particularly evident in EXP4, Figure 11, where the number of final social attitudes is almost 7.5 times the initial one. For highly fragmented societies, Figures12-13, the increase is slower due to the fact that the number of available independent social attitudes is reduced. For NC=1, Figures 11, the size distribution presents maximum for small values, and a decrease till size 1250. For highly fragmented societies, Figures 12-13,size distribution is peaked around small values of size and very narrow.

2) ε =1/5 The number of social attitudes increases faster than the case ε=0, (except in the case NC=100, Figure 13, where they coincide). Despite a better chance for merging and consequently reducing the number of different social attitudes, at this particular value, creation of new social attitude is enhanced. For NC=1, size distribution is wide and presents a maximum around 250. For larger value of NC, the size distribution is almost coincident with the previous case

3) ε >1/5 Independently of the initial fragmentation, the number of social attitudes reaches the value 2.5. Figures 11-13, right plots, show the existence of a large social attitude shared by almost 75% of the agents and a certain number of small social attitudes. Consequently the size distribution is the same independently of initial fragmentation whose only effect is increasing the convergence time (i.e. time to reach consensus).

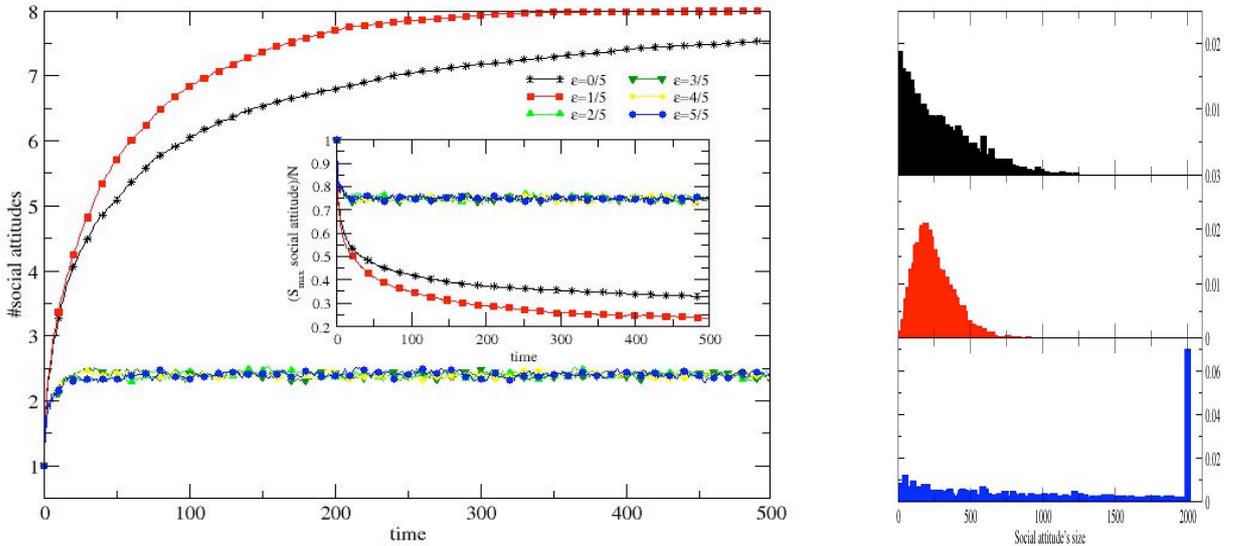

**Figure 11:** Left plot: The number of independent social attitudes as a function of time for different values of ε in the case of EXP4. Inbox: Normalized size of the biggest social attitudes as a function of time. Right plot: Size distribution of the social attitudes at the final state, for ε=0,1/5, 5/5. The results are averaged over 500 realizations of the system.

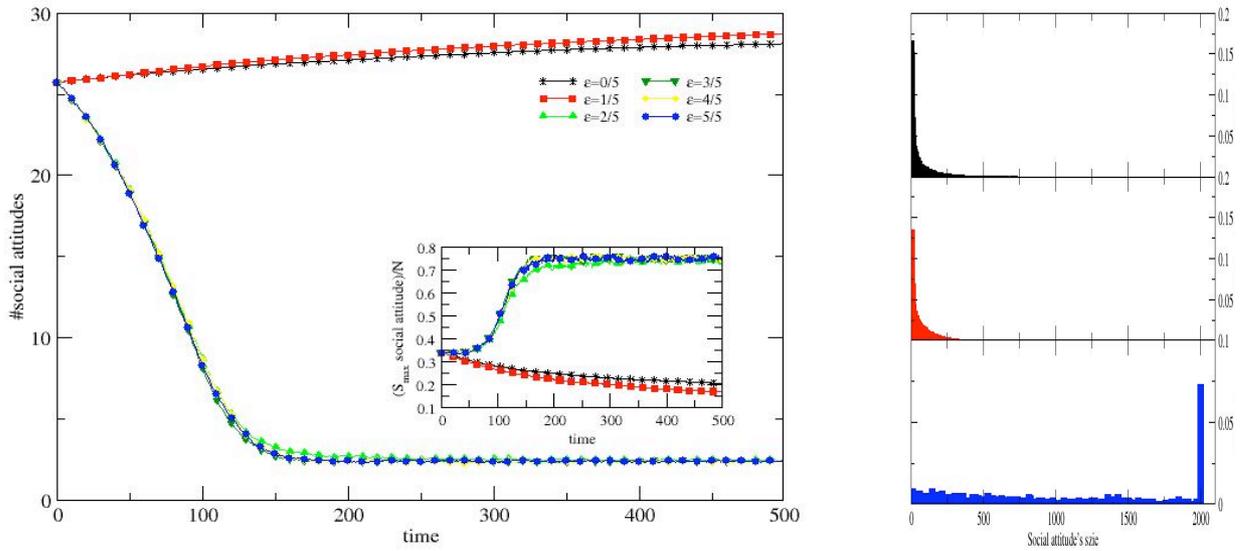

**Figure 12:** Left plot: The number of independent social attitudes as a function of time for different values of ε in the case of EXP5. Inbox: Normalized size of the biggest social attitudes as a function of time. Right plot: Size distribution of the social attitudes at the final state, for ε=0,1/5, 5/5. The results are averaged over 500 realizations of the system.

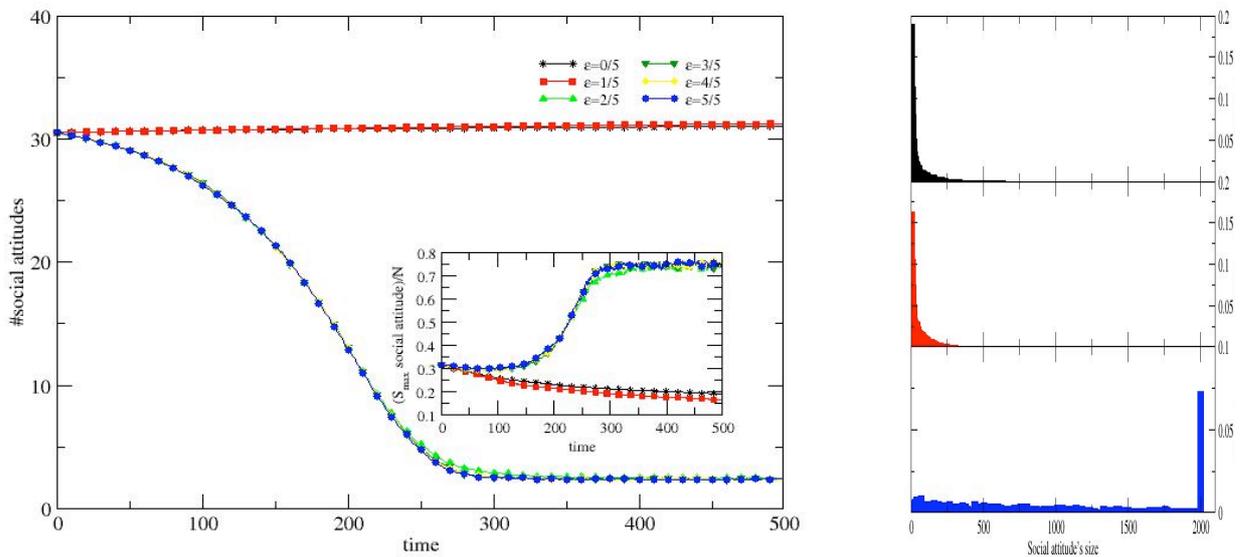

**Figure 13:** Left plot: The number of independent social attitudes as a function of time for different values of ε in the case of EXP6. Inbox: Normalized size of the biggest social attitudes as a function of time. Right plot: Size distribution of the social attitudes at the final state, for ε=0,1/5, 5/5. The results are averaged over 500 realizations of the system.

Figures 14-16 show the effects of the fragmentation probability, α, on the final number of independent social attitudes, expressed as a fraction of the initial number, for three critical values of

ε=1/5,2/5,5/5. We notice that independently of α for ε=1/5,2/5 , highly fragmented societies (respectively NC>50 and NC>20) have the same final number of social attitudes. Due to the high level of initial fragmentation and, then, the large number of independent social attitudes, the probability of creating new ones is almost null also for α=1. So the splitting process influences the number of groups and not the number of social attitudes. The net effect is a re-distribution of agents in the different social attitudes. The different values of ε speed up the process of reduction of social attitudes.

This conclusion cannot be extended to ε=5/5, Figures 16. In this case final numbers of social attitudes are different for every value of NC: while in a dynamical society the splitting process is countervailed by the merging one, in the static case groups can always and only merge consequently reducing the number of social attitudes in a faster way than the dynamical case.

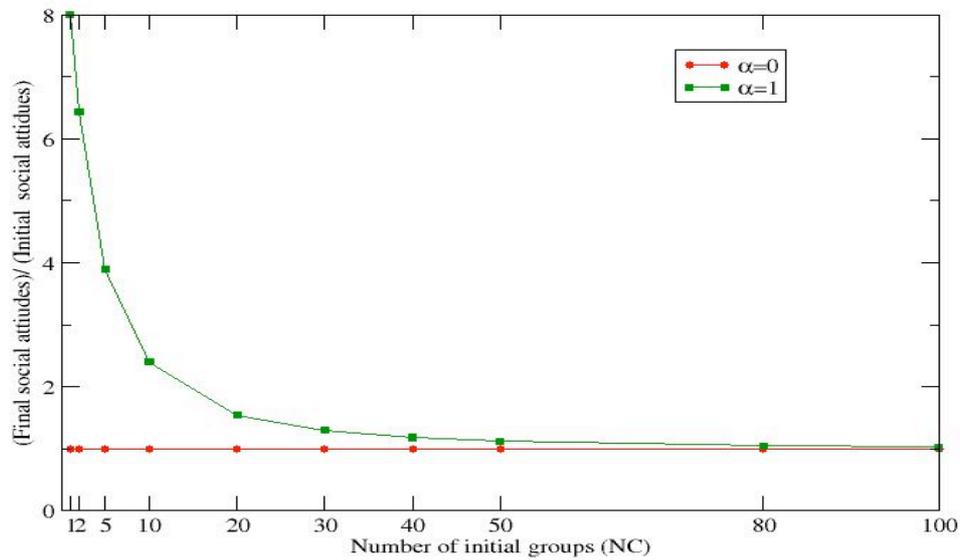

**Figure 14:** Final number of social attitudes divided by initial one as a function of the number of initial groups for ε=1/5.

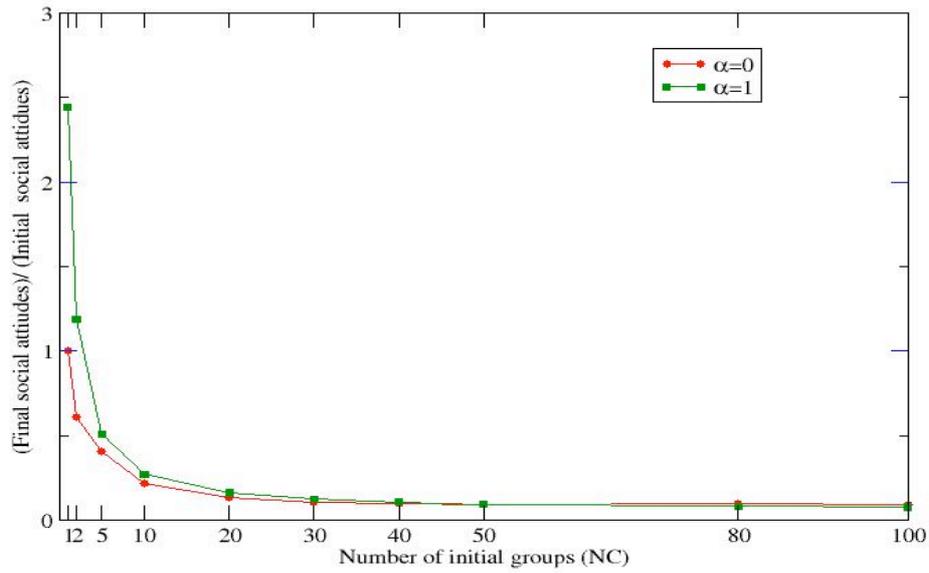

**Figure 15:** Final number of social attitudes divided by initial one as a function of the number of initial groups for ε=2/5.

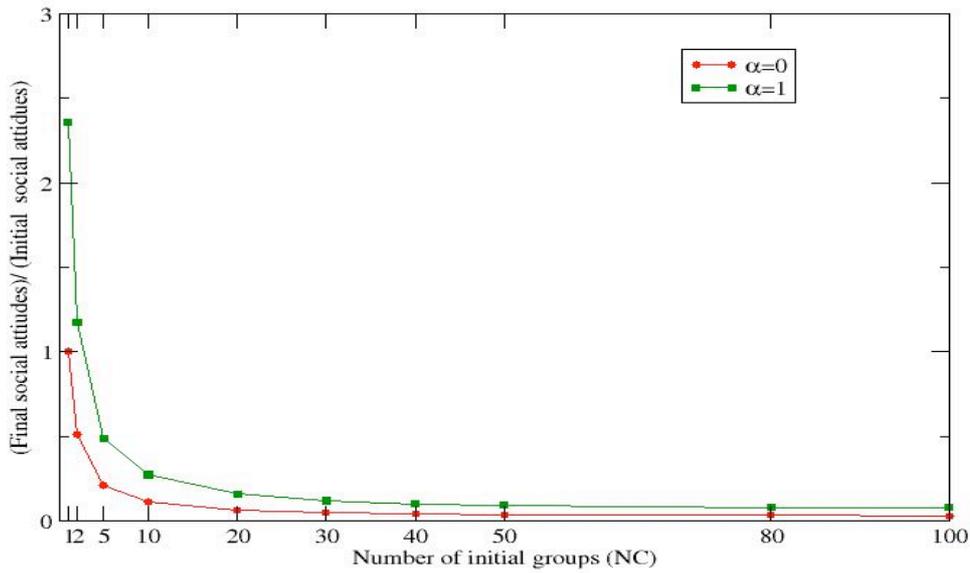

**Figure 16:** Final number of social attitudes divided by initial one as a function of the number of initial groups for ε=5/5.

# Diffusion of innovation

Figures 17-22 display the behaviour of innovation diffusion for the different sets of experiments. We analyze two main quantities: the percentage of innovators at each time step and the stabilization time of the process.

The measured quantities are averaged only on similar realizations: as in epidemics we consider outbreak realizations only in the cases where the innovative spreading leads to an increase in the initial number of innovators. The averages will be calculated only on such realizations.

The process starts with a single innovator located in a random community. The stabilization time is the time elapsed since the beginning of the process for reaching the equilibrium value between the two epidemic processes.

Figure 17 represents EXP1, the static case where, at each time step, a single community is present and, therefore, the population is a set of fully interacting agents.

The experiment reproduces the result of the discrete model described in the introduction: when the strength of the innovative idea is higher than the strength of the conservative one ($\beta_A$=0.6, $\beta_B$ =0.2), the innovation invade all the society, bringing to a new uniform point of view.

The stochasticity introduced through the binomial extraction reduces the percentage of outbreak realizations that, starting from one single innovator, results to be:

$$p_{outbreak}=74\%$$

Figures 18-19 display the results for EXP2-3. In both experiments, due to the absence of any type of dynamics in the case $\varepsilon$=0 (black lines), innovation is limited only to the group where the first innovator is seeded; due to the high initial fragmentation, just few agents can adopt the innovative idea. When the open-mindedness parameter $\varepsilon$=1/5, only groups with the same opinion can merge. This process allows the epidemics to spread to a larger portion of the population, with respect to the case $\varepsilon$=0, but at the same time creates barriers among groups that cannot be crossed, limiting the innovation process. For larger values of $\varepsilon$, the merging process is enhanced and the initial fragmentation facilitates the creation of a critical mass (the smaller is the group the easier is the contamination for the innovator) and then the self-sustainability of the process that in a short time involves all the population.

The initial fragmentation does not influence the possible equilibrium state, but only the time of the stabilization, as shown in the inset of Figures 18-19. Moreover, for a fixed value of the initial fragmentation, the stabilization time undergoes a regime transition depending on $\varepsilon$: for small values of $\varepsilon$ (0,1/5), the stabilization time increases with $\varepsilon$, while for higher values of $\varepsilon$ it is constant and independent of the final size of the contamination.

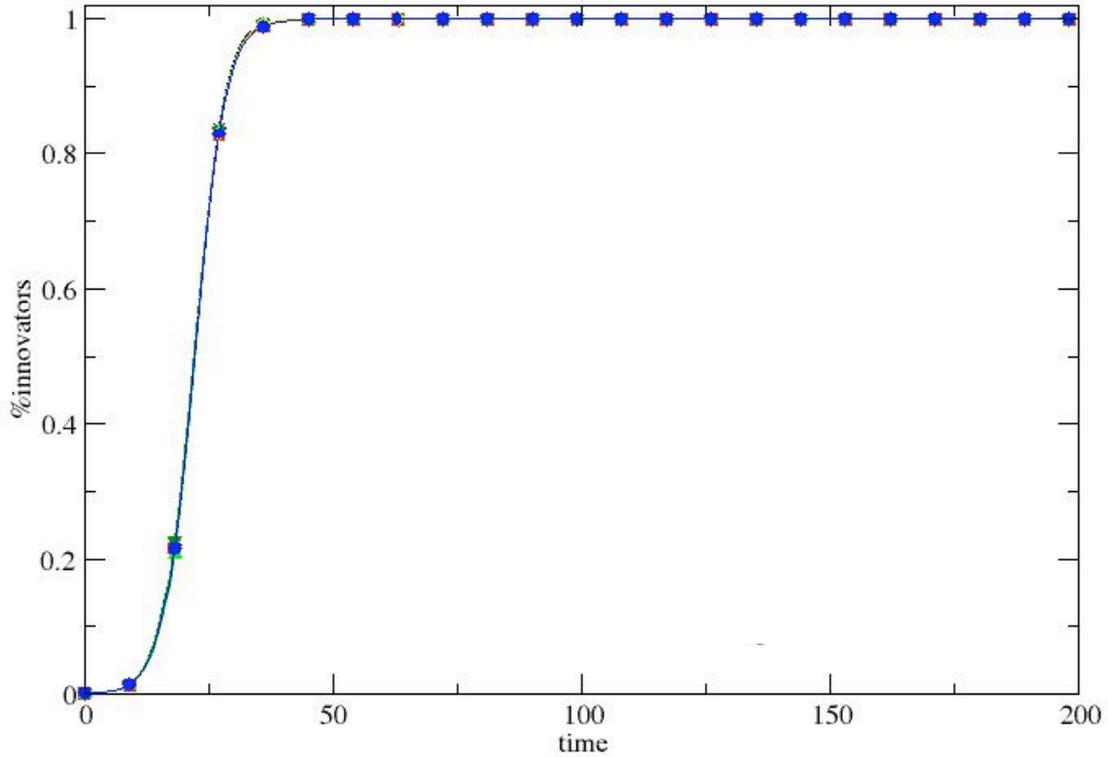

**Figure 17** Percentage of innovators as a function of time in the case of EXP1. The result is averaged on 100 outbreak realizations.

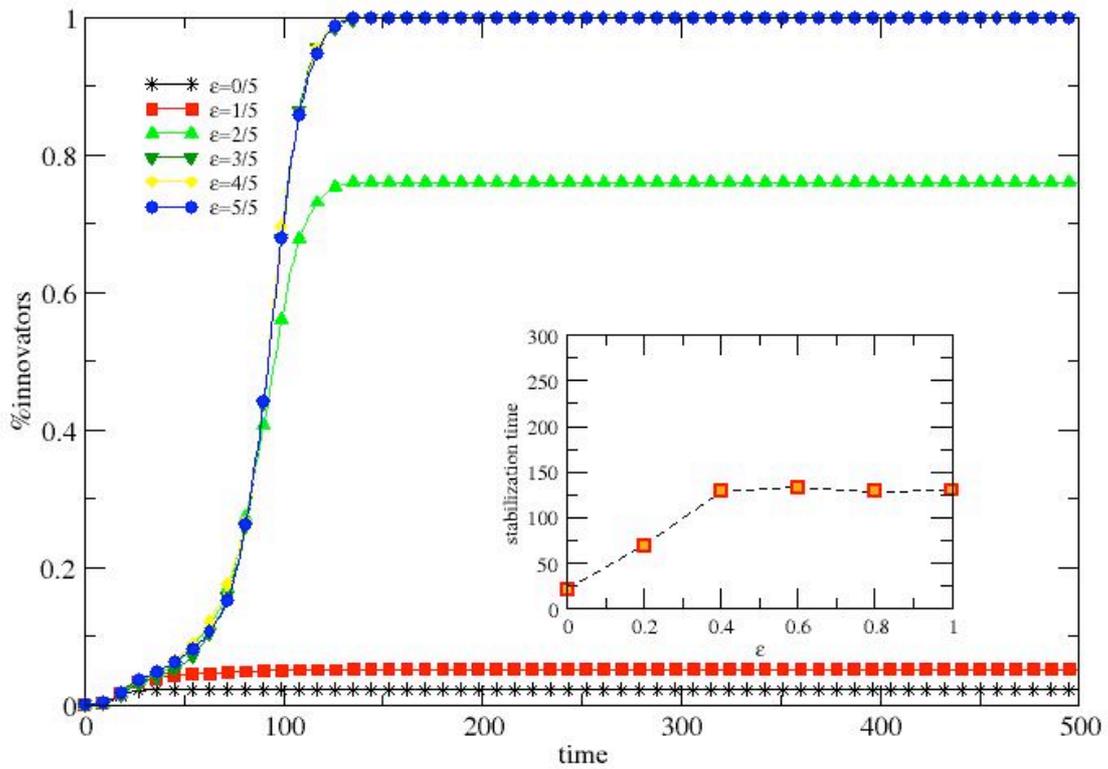

**Figure 18** Percentage of innovators as a function of time for different values of ε in the case of EXP2 Inbox: Stabilization time as a function of the open-mindedness parameter ε. The result is averaged on 100 outbreak realizations.

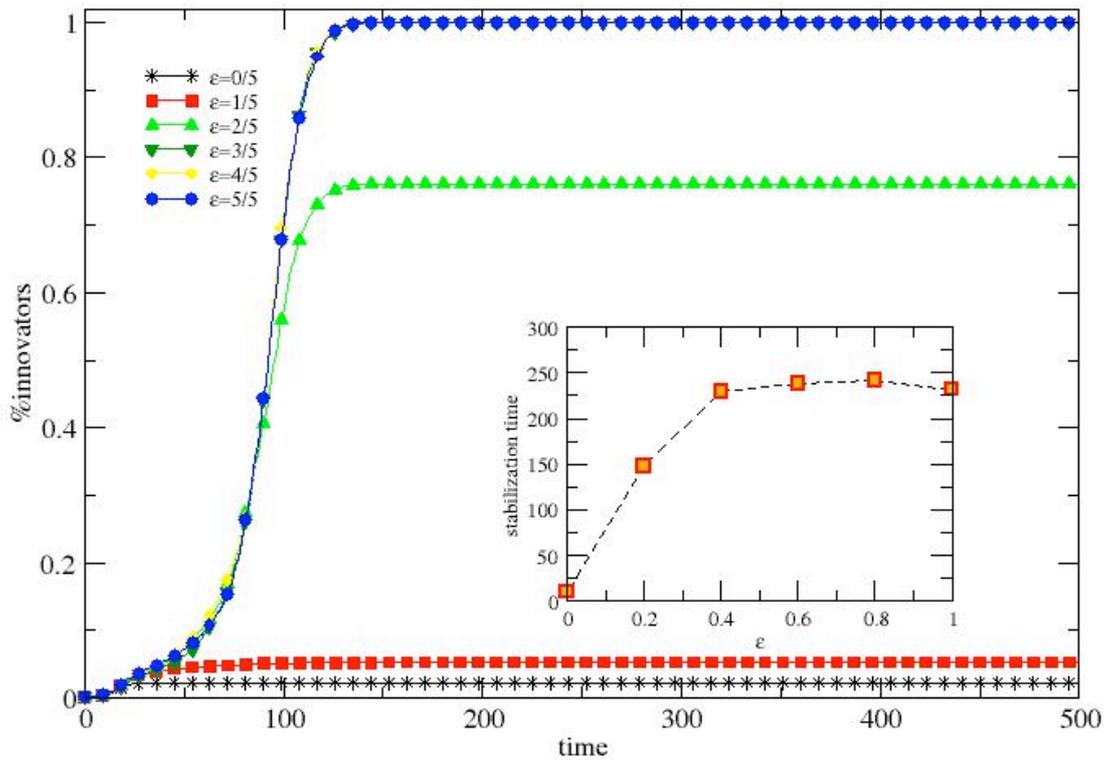

**Figure 19** Percentage of contaminated agents as a function of time for different values of ε in the case of EXP3 Inbox: Stabilization time as a function of the open-mindedness parameter ε. The result is averaged on 100 outbreak realizations.

The remaining experiment sets, EXP4, EXP5, EXP6 represent a dynamical society, where groups can coalesce and disaggregate.
As we can notice in Figure 20, referring to EXP5, dynamicity is an obstacle for the innovation process for NC=1. In particular, for ε=0, the fast splitting mechanism of the groups, not compensated form by a re-joining dynamics, suddenly creates a fragmented society where the communication is cut-off by strong cultural barriers. In this background, the innovative idea can reach only half of the population.
For higher values of ε, coalescence can occur, increasing the probability for the innovators to enter in contact with all the other agents, and consequently to transmit their idea.
In this case for each value of ε>0, the innovative idea spreads to all the population.
For ε=1/5, nevertheless, the stabilization time is higher, because of the slower grouping mechanism.

Figure 21 and 22 represent higher values of the initial fragmentation, namely EXP5, EXP6.
Comparing EXP5 and EXP6 with the static cases (respectively EXP2 and EXP3) we can see that, in this case, dynamicity gives in general a positive contribution to the innovation spreading: for ε=2/5 the innovation spreads to all the population while, for ε=1/5, it reaches a bit higher number of persons. But, even if the final result for the innovation spreading, for NC>1, is globally better in the dynamical cases, we should notice that the processes in the dynamical case are much slower.
Also for EXP4, EXP5 and EXP6, the stabilization time depends on the initial fragmentation: configurations starting from a smaller number of communities reach the equilibrium faster.

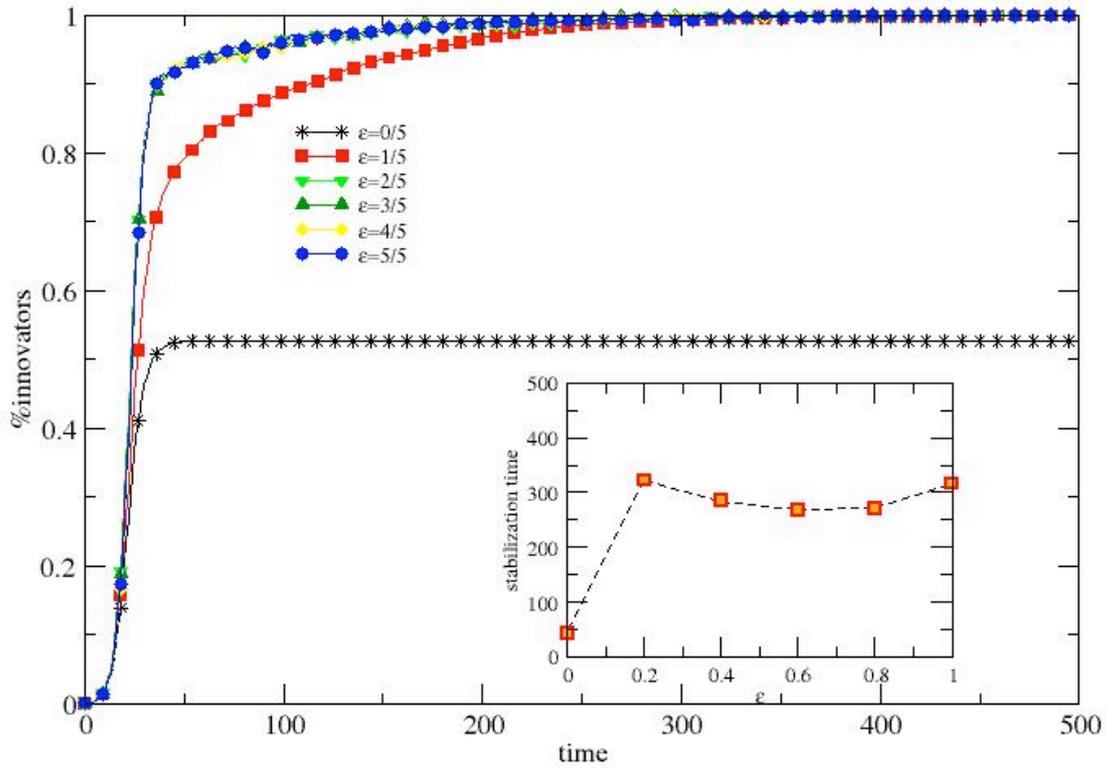

**Figure 20** Percentage of innovators as a function of time for different values of ε in the case of EXP4 Inbox: Stabilization time as a function of the open-mindedness parameter ε. The result is averaged on 100 outbreak realizations.

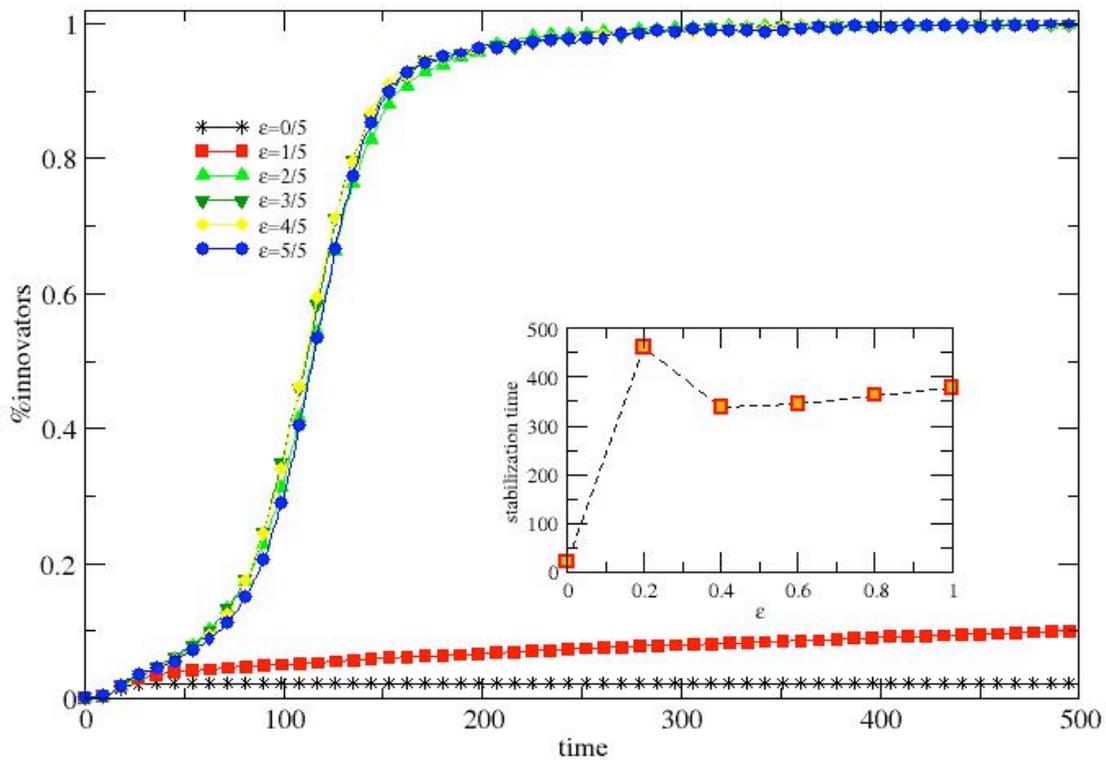

**Figure 21** Percentage of innovators as a function of time for different values of ε in the case of EXP5 Inbox: Stabilization time as a function of the open-mindedness parameter ε. The result is averaged on 100 outbreak realizations.

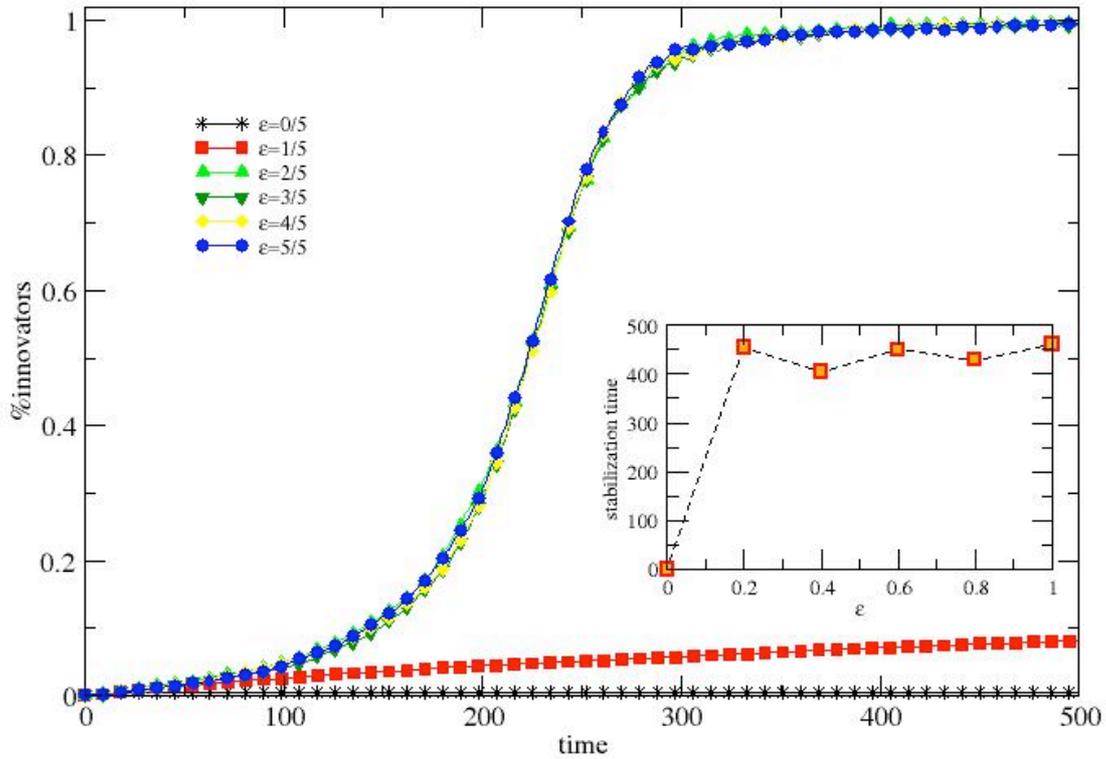

**Figure 22** Percentage of innovators as a function of time for different values of ε in the case of EXP6 Inbox: Stabilization time as a function of the open-mindedness parameter ε. The result is averaged on 100 outbreak realizations.

## CONCLUSIONS

In this article we have considered three social processes: groups' dynamics, majority formation and diffusion of innovation in a static and a dynamical society.
These three processes depend on a very large set of parameters and, in this paper, the attention has been focused on 3 of them: the initial number of groups (NC), the probability of splitting α and the open-mindedness parameter ε. In particular, we distinguish between a static society, where new social attitudes cannot be originated corresponding to the case α=0, and a dynamical one, corresponding to the case α=1, where new groups can be formed at each moment of the simulation.
For each of these two cases we have considered three different initial fragmentations of the society:
- A non-fragmented society corresponding to NC=1: all the agents belong to the same group, and initially there is just one social attitude. In the case of a static society, due to the absence of the fragmentation process, the society remains in this state for all the time.
- A medium fragmented society corresponding to NC=50.
- A highly fragmented society corresponding to NC=100: all the possible social attitudes are present in the society.

For all the possible types of societies the coalescence dynamic is ruled by the open-mindedness parameter ε. We considered all the possible values ranging from a close-minded society (ε=0) to an extremely open-minded one (ε=1).

This paper analyzes the time evolution of groups, social attitude and diffusion of innovations, in various zones of the parameter space.

In both static and dynamical society we could distinguish three phases for groups and social attitudes dynamics depending on ε. The specific behaviour according to ε depends on the kind of society (i.e. α).

For a static society, the phases are defined by ε=1/5 (phase 1), ε=2/5 (phase 2) and ε>2/5 (phase 3).

Phase 1: the number of groups is reduced to almost 50% of the original one. The size distribution at the end of the process reveals the majority of small groups and increases with NC.

In this phase the number of different social attitudes remains constant for all the duration of the process: agents redistribute among the different social attitudes and the final size distribution shows an abundance of small–sized social attitudes. In this phase a clear majority attitude is not realized.

Phase 2: at the end of the process society is divided into a small number of groups. A giant group is always present with few other small subgroups belonging to different social attitudes. In this case a majority attitude is always present.

Phase 3: consensus is reached and agents belong to just one group and consequently there is just one social attitude.

For a dynamical society the phases are defined by ε=0 (phase 1), ε=1/5 (phase 2) and ε>1/5 (phase 3).

Phase 1: independently of the initial fragmentation the number of groups and the number of social attitudes increase.

Phase 2: at the end of the process the society is highly fragmented and the number of social attitudes is increased by an amount depending on the initial configuration.

Phase 3: corresponds to phase 2 in the case of a static society.

We notice that for large values of NC (highly fragmented societies) and ε<= 2/5, the features of two kinds of society tend to become more similar. This effect is mainly due to the fact that in these cases almost all the possible social attitudes are present enhancing the process of coalescence.

When dealing with diffusion of innovation process the scenario is wider.

In the absence of group dynamics the process evolves according to the logistic map, bringing to the complete invasion of the scenario by the innovative idea.

Again we distinguish between static and dynamical society.

- Static society: the innovation size does not depend on initial number of groups. As in the case of social attitude we can distinguish three regimes for the final size of contamination: ε <=1/5, the contamination involves a small fraction of the population; ε=2/5 contamination happens involving almost 80% of the population; ε>2/5 contamination involves all the agents.
- Dynamical society: initial fragmentation strongly affects the process, giving different results for highly fragmented societies and unique group societies.
  Starting from a unique group and for ε=0, the innovation process is quenched and the final size of innovators attests around 50% of the population. For other values of ε, the only effect of the dynamics with respect to the static case is a general slowing down of the process.
  For highly fragmented societies, diffusion of the innovative idea is strongly suppressed not only for ε=0, but also for ε=1/5: when the coalescence process can not compensate for the splitting mechanism, the innovative ideas are destined to be supported only by a minority.

## ACKNOWLEDGMENT

We thank our external collaborators and members of the Network Dynamics and Simulation Science Laboratory (NDSSL) for their suggestions and comments. This work has been partially supported NSF Nets Grant CNS- 0626964, NSF HSD Grant SES-0729441, CDC Center of Excellence in Public Health Informatics Grant 2506055-01, NIH-NIGMS MIDAS project 5 U01 GM070694-05, DTRA CNIMS Grant HDTRA1-07-C-0113 and NSF NETS CNS- 0831633.